\documentclass{aa}

\usepackage{graphicx}
\usepackage{txfonts}
\usepackage[draft]{hyperref}
\usepackage{savesym}
\usepackage{txfonts}
\usepackage{amsmath}
\usepackage{aas_macros}
\usepackage{pdflscape}
\usepackage{afterpage}

\usepackage{color}
\definecolor{deepblue}{rgb}{0,0,0.5}
\definecolor{deepred}{rgb}{0.6,0,0}
\definecolor{deepgreen}{rgb}{0,0.5,0}
\definecolor{mygrey}{rgb}{0.5,0.5,0.5}

\usepackage{natbib,twoopt}
\bibpunct{(}{)}{;}{a}{}{,} 

\makeatletter
\newcommandtwoopt{\citeads}[3][][]{\href{http://adsabs.harvard.edu/abs/#3}%
{\def\hyper@linkstart##1##2{}%
\let\hyper@linkend\@empty\citealp[#1][#2]{#3}}}
\newcommandtwoopt{\citepads}[3][][]{\href{http://adsabs.harvard.edu/abs/#3}%
{\def\hyper@linkstart##1##2{}%
\let\hyper@linkend\@empty\citep[#1][#2]{#3}}}
\newcommandtwoopt{\citetads}[3][][]{\href{http://adsabs.harvard.edu/abs/#3}%
{\def\hyper@linkstart##1##2{}%
\let\hyper@linkend\@empty\citet[#1][#2]{#3}}}
\newcommandtwoopt{\citeyearads}[3][][]%
{\href{http://adsabs.harvard.edu/abs/#3}
{\def\hyper@linkstart##1##2{}%
\let\hyper@linkend\@empty\citeyear[#1][#2]{#3}}}
\makeatother

\begin{document}
\title{HI4PI: A full-sky \ion{H}{i} survey based on EBHIS and GASS}
\author{\textit{HI4PI Collaboration:}\\
    N.~Ben~Bekhti\inst{1}, L.~Fl\"{o}er\inst{1}, R.~Keller\inst{2}, J.~Kerp\inst{1}, D.~Lenz\inst{1}, and B.~Winkel\thanks{Corresponding authors:~B.~Winkel (\email{bwinkel@mpifr.de}) and P.\,M.\,W.~Kalberla (\email{kalb@astro.uni-bonn.de})}\inst{1,2}\\
    \and\\
    J.~Bailin\inst{3,4}, M.\,R.~Calabretta\inst{5}, L.~Dedes\inst{1}, H.\,A.~Ford\inst{6}, B.~K.~Gibson\inst{7}, U.~Haud\inst{8}, S.~Janowiecki\inst{9}, P.\,M.\,W.~Kalberla$^\star$\inst{1}, F.\,J.~Lockman\inst{4}, N.\,M.~McClure-Griffiths\inst{10,5}, T.~Murphy\inst{11}, H.~Nakanishi\inst{12,13,14}, D.\,J.~Pisano\inst{15}, and L.~Staveley-Smith\inst{9,16}
   }

\institute{
    Argelander-Institut f\"{u}r Astronomie (AIfA), University of Bonn, Auf dem H\"{u}gel~71, 53121 Bonn, Germany
    \and
    Max-Planck-Institut f\"{u}r Radioastronomie (MPIfR), Auf dem H\"{u}gel~69, 53121 Bonn, Germany\\
    \and
    Department of Physics and Astronomy, University of Alabama, Box 870324, Tuscaloosa, AL, 35487-0324, USA
    \and
    National Radio Astronomy Observatory, P.\,O. Box 2, Green Bank, WV, 24944, USA
    \and
    CSIRO Astronomy and Space Science, PO Box 76, Epping, NSW 1710, Australia
    \and
    Steward Observatory, University of Arizona, Tucson, AZ 85721, USA
    \and
    E.\,A. Milne Centre for Astrophysics, University of Hull, Hull, HU6\,7RX, UK
    \and
    Tartu Observatory, 61602 T\~{o}ravere, Tartumaa, Estonia
    \and
    International Centre for Radio Astronomy Research (ICRAR), University of Western Australia, 35 Stirling Highway, Crawley, WA 6009, Australia
    \and
    Research School of Astronomy and Astrophysics, Australian National University, Canberra, ACT 2611, Australia
    \and
    Sydney Institute for Astronomy, School of Physics, University of Sydney, NSW, 2006
    \and
    Graduate School of Science and Engineering, Kagoshima university, 1-21-35 Korimoto, Kagoshima
    \and
    Institute of Space and Astronautical Science, Japan Aerospace Exploring Agency, 3-1-1 Yoshinodai, Sagamihara, Kanagawa 252-5210, Japan
    \and
    SKA Organization, Jodrell Bank Observatory, Lower Withington, Macclesfield, Cheshire SK11 9DL, UK
    \and
    Department of Physics and Astronomy, West Virginia University, P.\,O. Box 6315, Morgantown, WV 26506, USA
    \and
    ARC Centre of Excellence for All-sky Astrophysics (CAASTRO)
    }

\date{Received ; accepted }

\abstract
{Measurement of the Galactic neutral atomic hydrogen (\ion{H}{i}) column density, $N_\ion{H}{i}$, and brightness temperatures, $T_\mathrm{B}$, is of high scientific value for a broad range of astrophysical disciplines. In the past two decades, one of the most-used legacy \ion{H}{i} datasets has been the Leiden/Argentine/Bonn Survey (LAB).}
{We release the \ion{H}{i} $4\pi$ survey (HI4PI), an all-sky database of Galactic \ion{H}{i}, which supersedes the LAB survey.}
{The HI4PI survey is based on data from the recently completed first coverage of the Effelsberg--Bonn \ion{H}{i} Survey (EBHIS) and from the third revision of the Galactic All-Sky Survey (GASS). EBHIS and GASS share similar angular resolution and match well in sensitivity. Combined, they are ideally suited to be a successor to LAB.}
{The new HI4PI survey outperforms the LAB in angular resolution ($\vartheta_\mathrm{fwhm}=16\farcm2$) and sensitivity ($\sigma_\mathrm{rms}=43~\mathrm{mK}$). Moreover, it has full spatial sampling and thus overcomes a major drawback of LAB, which severely undersamples the sky. We publish all-sky column density maps of the neutral atomic hydrogen in the Milky Way, along with full spectroscopic data, in several map projections including HEALPix.\thanks{HI4PI datasets are available in electronic form (FITS) at the CDS via anonymous ftp to cdsarc.u-strasbg.fr (130.79.128.5) or via\newline\url{http://cdsweb.u-strasbg.fr/cgi-bin/qcat?J/A+A/594/A116}}}
{}

\titlerunning{HI4PI: A full-sky \ion{H}{i} survey based on EBHIS and GASS}
\authorrunning{HI4PI collaboration}

\maketitle

\section{Introduction}\label{sec:intro}

Hydrogen is the most abundant element in space and as such it is a significant tracer for gas in the Universe \citep{burbidge57}. Because atomic neutral hydrogen (\ion{H}{i}) is considered to trace Galactic interstellar matter quantitatively, its distribution is commonly used to supplement observations from the gamma-ray to the far-infrared regime, for example, to extract the cosmic infrared background (CIB) from Planck satellite data \citep[][and references therein]{planck14}. The 21-cm hyperfine structure line of \ion{H}{i} was predicted by \citet{vandehulst45}. With radio observations it can easily be observed in emission and absorption \citep[e.g.,][]{ewen51,radhakrishnan72,dickey83,dickey03}. Soon after the discovery of the 21-cm line \citep{muller51,ewen51}, surveys of the Milky Way (MW) \ion{H}{i} distribution were undertaken with great success. For example, analyzing the \ion{H}{i} data provided independent evidence of the MW being a spiral galaxy \citep{vandehulst54,oort58}. The early survey observations were made with relatively small dishes of diameter less than 10~m. A few years later, \citet{weaver73} used the 85-foot Hat Creek telescope for a Galactic plane survey ($\vert b\vert\leq10\degr$), the Berkeley Low-Latitude Survey of Neutral Hydrogen, which was subsequently complemented by a survey for $\vert b\vert\geq10\degr$ (north of declination $\geq$$-30\degr$) using the same instrument \citep{heiles74}. The Bell Laboratories \ion{H}{i} survey \citep[declination $\geq$$-40\degr$;][]{stark92} is another example of a sensitive large-area survey. It also offered lower side-lobe contamination compared to earlier datasets, but with only limited spectral resolution ($\delta\varv=5.3~\mathrm{km\,s}^{-1}$).

In the past two decades, the Leiden/Argentine/Bonn Survey \citep[LAB;][]{bajaja85,kalberla05} has been the prime source of information on \ion{H}{i}. LAB was merged using the Leiden/Dwingeloo Survey \citep[LDS;][]{burton94,hartmann97} and the Instituto Argentino de Radioastronom\'{ı}a Survey \citep[IAR;][]{arnal00,bajaja05}, both made with 25-m class telescopes. LAB was the first all-sky \ion{H}{i} survey corrected for stray radiation \citep[SR;][]{kalberla80a,kalberla80b}. SR is the result of emission collected via the side-lobes of a radio antenna. Toward high galactic latitudes, where $N_\ion{H}{i}$ is low, the SR contribution can even exceed the true hydrogen column density. This is because a large fraction of the far side-lobe pattern can be directed at the luminous Milky Way disk in such a case.

Most recently, the GALFA-HI survey \citep{peek11} was initiated exploiting the huge 300-m Arecibo dish and state-of-the-art spectrometers ($\delta\varv=0.2~\mathrm{km\,s}^{-1}$). While it excels in sensitivity, the Arecibo telescope is not fully steerable and the fraction of the sky it can access is therefore limited. Furthermore, there is currently no SR correction software available for \ion{H}{i} observations made with the 300-m dish.

Galactic \ion{H}{i} is so ubiquitous that in fact no single sight line exists through the Milky Way where it could not be detected after only a few seconds of integration with a modern radio telescope. Nonetheless it took a long time for the largest fully steerable radio telescopes in the world to begin the endeavor of imaging the full sky in the 21-cm line. Only with the advent of multi-beam receivers \citep{staveley96,klein04,keller06} about two decades ago, was it finally feasible to conduct all-sky surveys, which still required thousands of hours of observing time. Another important aspect was the availability of cheap correlator technology to process the multi-beam signals.

Two such surveys, the Effelsberg--Bonn \ion{H}{i} Survey \citep[EBHIS;][]{kerp11,winkel16a} and the Galactic All-Sky Survey \citep[GASS, performed with the Parkes telescope;][]{mcclure09,kalberla10,kalberla15}, mapped the complete northern and southern hemisphere, respectively. Full spectroscopic data of the EBHIS were recently released, while GASS has been available to the scientific community since 2009 \citep{mcclure09}, with revisions published in 2010 \citep{kalberla10} and 2015 \citep{kalberla15}. There is a noteworthy difference between EBHIS and GASS: the Effelsberg spectroscopy back-ends allowed the mapping of not only the Galactic velocity regime but also the extra-galactic sky out to a redshift of $z\sim0.07$. For the Southern sky there is a separate extra-galactic \ion{H}{i} survey, the \ion{H}{i} Parkes All-Sky Survey \citep[HIPASS,][]{barnes01}.

The EBHIS and GASS Milky Way data provide an excellent database to approach a wealth of scientific questions. In the past it has been used to study the MW halo \citep{ford08,winkel11,benbekhti12,venzmer12,hernandez13,moss13,for14,roehser14,hammer15,lenz16,kerp16b} and the disk--halo interaction \citep{ford10,mcclure10,lenz15,roehser16a}. Likewise the MW disk material itself can be explored in much greater detail than previously feasible \citep[e.g.,][]{haud13,kalberla16a,kalberla16b}, also revealing spectacular Galactic super-shells \citep{mcclure06,moss12}. Most recently, \citet{kerp16a} reported the existence of a giant halo of \ion{H}{i} around the Andromeda galaxy (\object{M\,31}).

We now aim to substitute the all-sky LAB survey using the latest data from the EBHIS and GASS, both offering much higher angular resolution and sensitivity. Furthermore, LAB sampled the sky on a beam-by-beam grid only, which leads to difficulties in reconstructing small-scale features \citep[see e.g.,][]{kerp11,kalberla16a}. The result of the merged EBHIS and GASS datasets is hereafter called the HI4PI Survey.

Since 2010, when the work on the third version of GASS began, EBHIS and GASS data reduction software was developed in close collaboration, with the ultimate goal of merging the two datasets. It turned out to be very beneficial that the software development for both surveys was done at the same site (Bonn), because intermediary and final data products were kept compatible early on. For example, a common HEALPix grid \citep{gorski05} was used for internal storage of spectra and we took care to use a common brightness temperature scale for both surveys.

In Section~\ref{sec:observations} we summarize the observing strategies and data processing for both surveys. The resulting full-sky \ion{H}{i} column density map, $N_\ion{H}{i}$, is presented in Section~\ref{sec:momentmaps}, integrated over the full HI4PI velocity interval. Thus the \ion{H}{i} also comprises extra-planar objects (e.g., intermediate- and high-velocity clouds) and nearby galaxies, which leads to a certain contamination of the Galactic $N_\ion{H}{i}$ map. This is discussed in Section~\ref{subsec:contamination}.  Section~\ref{sec:dataproducts} describes how the new \ion{H}{i} column density maps and data cubes can be retrieved. We conclude with a summary in Section~\ref{sec:summary}. Furthermore, a comparison of EBHIS and GASS data quality and calibration consistency is made by inspecting the overlap area (declination range: $-5\degr\leq\delta\leq0\degr$) of the two surveys. This is presented in Appendix~\ref{sec:ebhis_gass_comparison}.

\section{Observations and data processing}\label{sec:observations}

\subsection{GASS observations}
GASS observations were performed from January 2005 to November 2006, utilizing the 13-beam 21-cm feed array installed at the Parkes 64-m telescope \citep{mcclure09}. To avoid solar interference, which can cause ripples in the \ion{H}{i} spectra, the measurements were carried out only during night time. Each area of the sky was covered twice, scanning in right ascension and declination, respectively. The feed array was rotated by $19\fdg1$ with respect to the scan direction to achieve a homogeneous distribution of the scan tracks. Data were recorded using a 2048-channel auto-correlator in in-band frequency-switching mode, storing one spectrum per feed and polarization every 5~s. The scan speed was $1\degr\,\mathrm{min}^{-1}$. The total bandwidth is 8~MHz which corresponds to a radial velocity coverage of $-470\leq\varv_\mathrm{lsr}\leq470~\mathrm{km\,s}^{-1}$.

\subsection{EBHIS observations}

EBHIS was conducted with the Effelsberg 100-m telescope, utilizing the seven-beam 21-cm receiver. Measurements were carried out between end of 2008 and early 2013. State-of-the-art fast Fourier transform spectrometers \citep[FFTS;][]{stanko05,klein12}, allowed us to simultaneously cover 100~MHz of bandwidth with 16\,384 spectral channels. The resulting frequency and velocity channel width is slightly larger than for GASS. Owing to the large bandwidth, both the Galactic and extra-galactic sky could be observed simultaneously. For the first data release of EBHIS \citep{winkel16a} the Galactic velocity regime ($-600\leq\varv_\mathrm{lsr}\leq600~\mathrm{km\,s}^{-1}$) was published. The FFTS can record spectra with integration (dump) times of $\sim100~\mathrm{ms}$. To reduce the raw data volume, \ion{H}{i} spectra were stored every $500~\mathrm{ms}$, which is an order of magnitude faster than in the case of GASS. The short integration times are beneficial for RFI mitigation of burst-like events \citep[see][for details]{winkel16a}. EBHIS was scanned along right ascension (scan speed: $4\degr\,\mathrm{min}^{-1}$). Analogous to GASS, a $19\fdg1$ feed rotation angle was used to ensure homogeneous angular sampling.

\subsection{Data reduction}

Pre-processing of GASS data was performed using the \textit{Livedata} package \citep{barnes01}, but for radio-frequency interference (RFI) mitigation, flux calibration, baseline fitting, and SR correction, dedicated software was used. For details, we refer to the GASS data reduction papers \citep{mcclure09,kalberla10,kalberla15}.

In comparison with GASS, EBHIS was processed using entirely different data reduction software and techniques \citep[][and references therein]{winkel10,winkel16a}. It was decided to not utilize the frequency-switching technique during processing. This was done to improve brightness temperature calibration by accounting for its frequency dependence, which is not easily possible for frequency switching \citep[see][for details]{winkel12}. Also, the very different RFI environment and spectrometers made it necessary to develop a well-adapted RFI mitigation tool \citep{floeer10,winkel16a}. Baselines were fitted using 2-D polynomials instead of 1-D polynomials as in the case of GASS.

The absolute flux calibration for EBHIS and GASS is performed in the same way as for the LAB survey. It is based on the ideas developed by \citet{kalberla82} and uses IAU standard line calibrators \object{S\,6}, \object{S\,7}, and \object{S\,8}. This ensures a well-matching brightness temperature- and flux density calibration of all these data sets. Also the stray-radiation correction follows the same approach for all three surveys, originally developed by \citet{kalberla80a,kalberla80b} for the 100-m telescope at Effelsberg and later improved upon and adapted to different telescopes \citep{kalberla05,kalberla10,winkel16a}. For both surveys, a  convolution-based data gridding technique is applied. For EBHIS, the gridding software, cygrid\footnote{\url{https://github.com/bwinkel/cygrid}}, was recently published under open-source license and we refer to \citet{winkel16b} for more details on the underlying algorithm. Originally, for the first GASS data release, the \textit{Gridzilla} software, which is part of \textit{Livedata}, was used for gridding \citep{barnes01}. The second and third data releases also use a Gaussian-convolution based gridding software \citep{kalberla10} but, unlike cygrid and Gridzilla, it lacks full support of the FITS \citep{hanisch01} world coordinate system standard \citep[WCS;][]{greisen02,calabretta02}. WCS is widely used to define map projections in FITS images.

\subsection{Merging the two data sets: HI4PI}

\begin{table}
\caption{Comparison of basic parameters of selected \ion{H}{i} surveys of the Milky Way. The table quotes the declination range,  $\delta$, angular resolution $\vartheta_\mathrm{fwhm}$, velocity interval, $\varv_\mathrm{lsr}$, channel separation, $\Delta \varv$, spectral resolution, $\delta \varv$, and brightness temperature noise level, $\sigma_\mathrm{rms}$.
The bottom two rows quote theoretical $5\sigma$ detection limits (velocity-integrated intensity) integrating over a Gaussian profile of $20~\mathrm{km\,s}^{-1}$ line width (FWHM) for surface brightness and point sources, respectively. \textit{Adapted from \citet{winkel16a}.}}
\label{tab:survey_comparison}
\centering
\begin{tabular}{l c c c c l}
\hline\hline
\rule{0ex}{3ex} & LAB & GASS & EBHIS & HI4PI  & Unit\\
\hline
\rule{0ex}{3ex}$\delta$ & Full & $\leq1\degr$ & $\geq-5\degr$ & Full\\
$\vartheta_\mathrm{fwhm}$ & $36\arcmin$ & $16\farcm2$ & $10\farcm8$ & $16\farcm2$ \\
$\vert \varv_\mathrm{lsr}\vert$ & $\leq460^\mathrm{\dagger}$ & $\leq470$     & $\leq600$ & $\leq600^\ast$ & $\mathrm{km\,s}^{-1}$\\
$\Delta \varv$ & $1.03$ & $0.82$    & $1.29$ & $1.29$ & $\mathrm{km\,s}^{-1}$\\
$\delta \varv$ & $1.25$ & $1.00$    & $1.49$& $1.49$& $\mathrm{km\,s}^{-1}$\\
$\sigma_\mathrm{rms}$ & $80$ & $55^\mathrm{\star}$    & $90$ & $\sim$$43$ & $\mathrm{mK}$\\
\hline
\rule{0ex}{3ex}$N_\ion{H}{i}^\mathrm{lim}$ & $3.9$ & $2.5$    & $4.7$ & $\sim$$2.3$ & $10^{18}~\mathrm{cm}^{-2}$\\
\rule{0ex}{3ex}$S_\ion{H}{i}^\mathrm{lim}$ & $16.1$ & $2.1$    & $1.8$ & $\sim$$2.0$ & $\mathrm{Jy~km\,s}^{-1}$\\[0.5ex]
\hline
\multicolumn{6}{l}{\rule{0ex}{3ex}$^\dagger$\,Northern (LDS) part: $\varv_\mathrm{lsr}\leq 400~\mathrm{km\,s}^{-1}$}\\
\multicolumn{6}{l}{\rule{0ex}{2ex}$^\ast$\,Southern (GASS) part: $\vert \varv_\mathrm{lsr}\vert\leq 470~\mathrm{km\,s}^{-1}$}\\
\multicolumn{6}{l}{\rule{0ex}{2ex}$^\star$\,Re-scaled intensity calibration \citep[see][]{kalberla15}}
\end{tabular}
\end{table}

Important survey parameters of EBHIS and GASS were compiled in \citet[][their Table~1]{winkel16a}. For the reader's convenience, we reproduce it here and add a column for HI4PI (Table~\ref{tab:survey_comparison}). Nominally, EBHIS has 30$-$40\% higher brightness temperature noise than GASS. However, smoothing EBHIS to GASS angular resolution and decreasing the spectral resolution of GASS to match the one of EBHIS yields almost the same average noise level of $\sim$$43~\mathrm{mK}$ in the two independent products used for HI4PI. Therefore, HI4PI offers a convenient and coherent data base of \ion{H}{i}, which is homogeneous in all basic parameters across the two hemispheres.

We note that because the extra-galactic part of EBHIS is less sensitive than HIPASS, the EBHIS team is currently conducting observations for a second coverage of the sky, which will bring EBHIS and HIPASS to a very similar sensitivity level. Like GASS, EBHIS will then have two orthogonal scan directions.

\begin{figure*}[!p]
\centering%
\includegraphics[width=\textwidth,viewport=13 19 670 230,clip=]{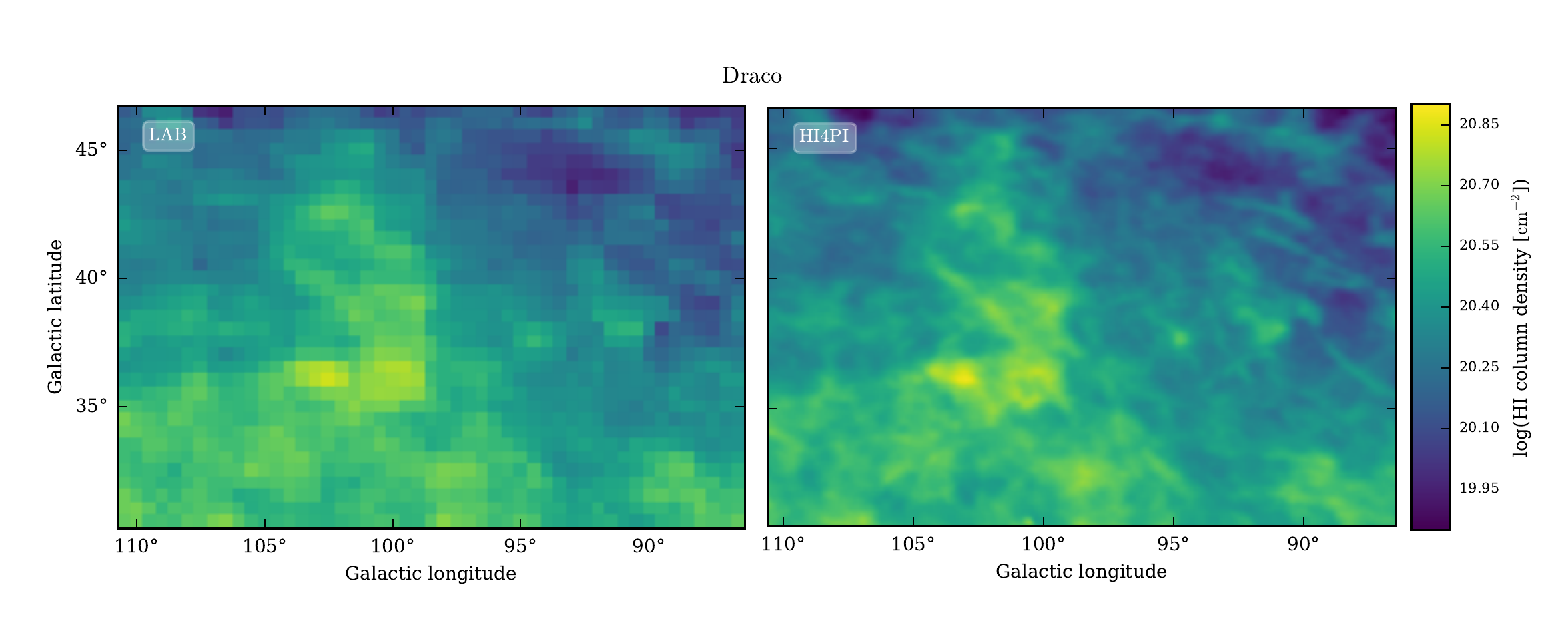}\\[0ex]
\includegraphics[width=\textwidth,viewport=13 7 670 243,clip=]{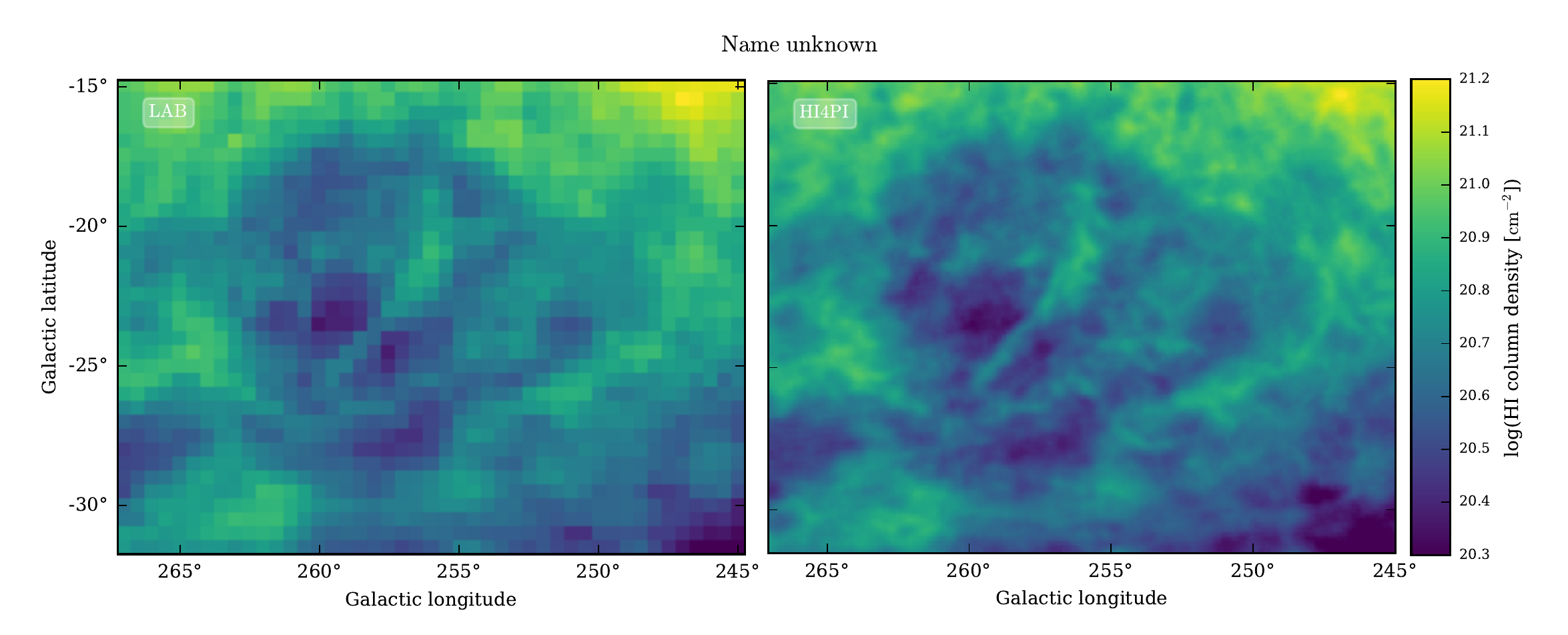}\\[0ex]
\includegraphics[width=\textwidth,viewport=13 28 670 222,clip=]{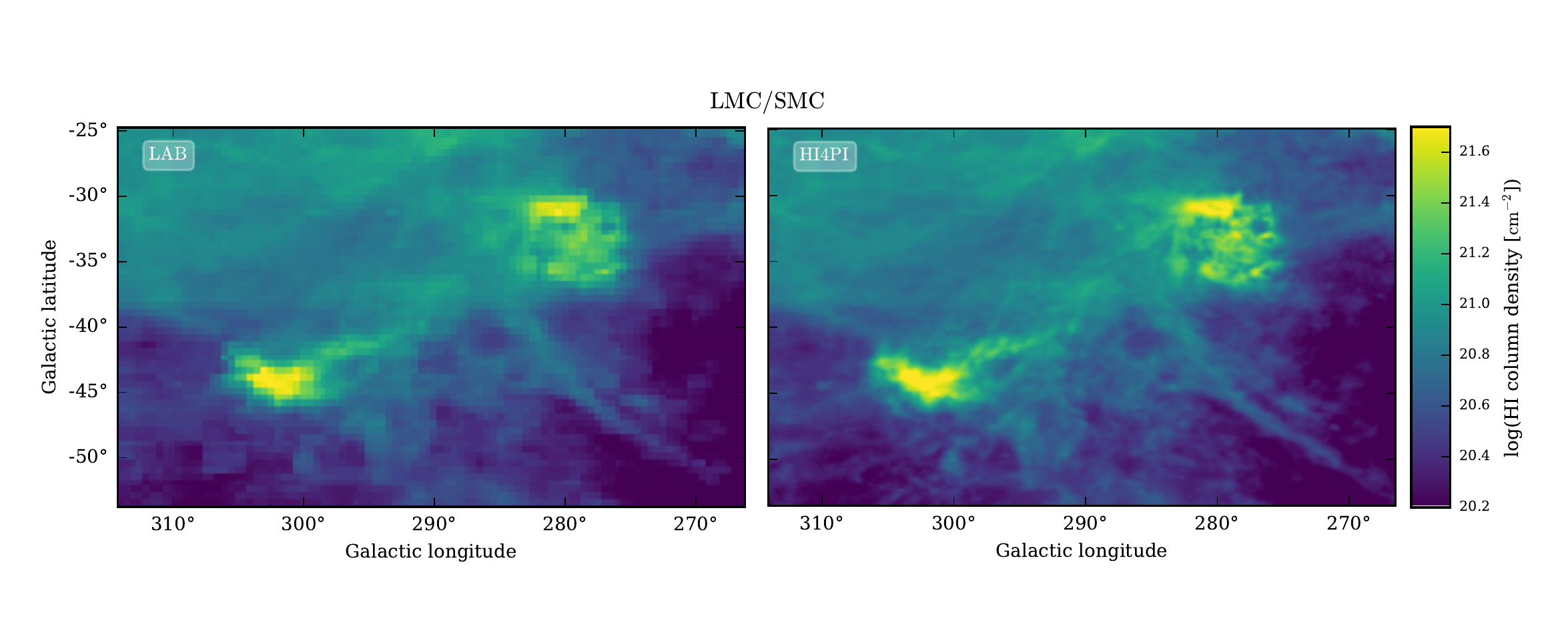}\\[0ex]
\includegraphics[width=\textwidth,viewport=13 22 670 226,clip=]{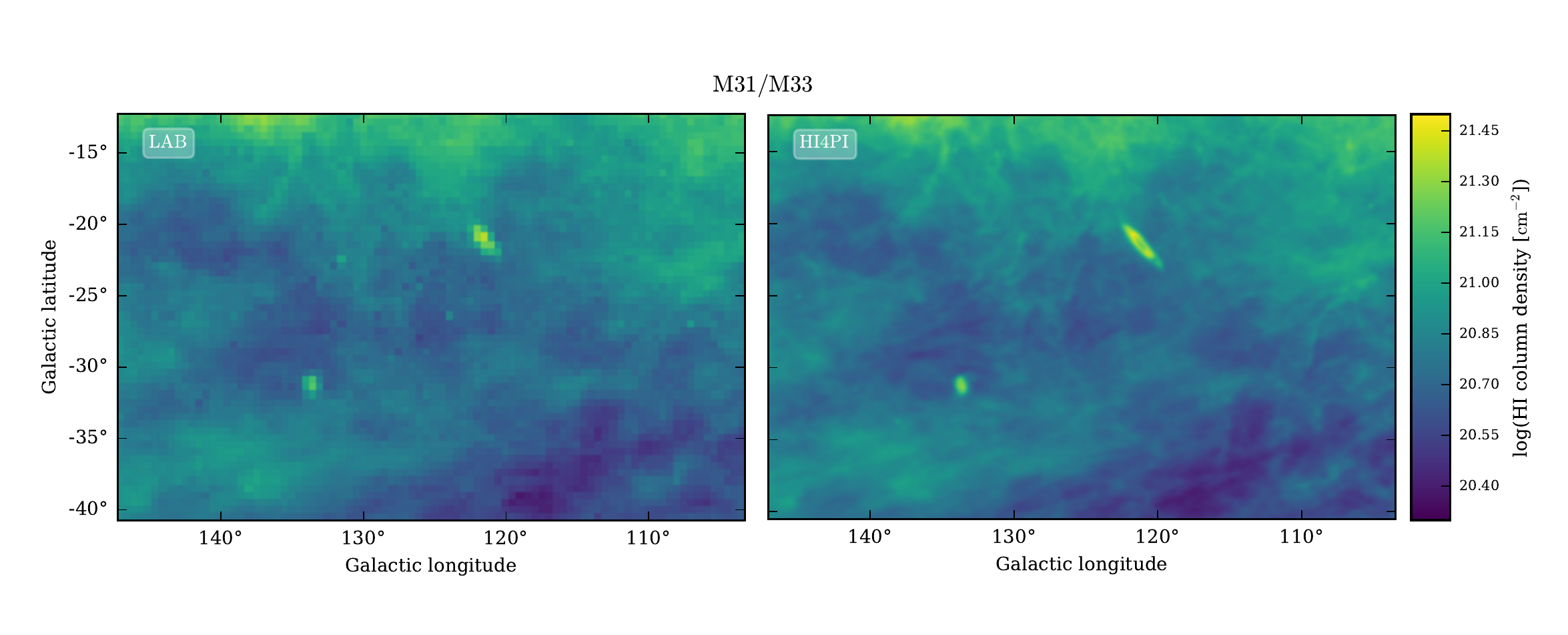}
\caption{Comparison between LAB and HI4PI column densities for selected regions, which are further discussed in Section~\ref{subsec:contamination}. The top row shows the Draco cloud, the second and the third row contain a part of the Leading Arm of the Magellanic Cloud system, including the \object{SMC} and \object{LMC}, respectively. The bottom row displays the environment of \object{M\,31} and \object{M\,33}.}%
\label{fig:hi4pi_lab_comparison}%
\end{figure*}

In \citet{winkel16a} it was demonstrated that EBHIS and GASS yield very consistent column densities and brightness temperatures (see also Appendix~\ref{sec:ebhis_gass_comparison}). With respect to the LAB survey no significant bias in the intensity scales could be identified, implying that the new HI4PI dataset can safely be used as a drop-in replacement for the former LAB data. To demonstrate the improvement in image fidelity of HI4PI compared to the LAB survey, we show in Fig.~\ref{fig:hi4pi_lab_comparison} $N_\ion{H}{i}$ maps of four regions containing very distinct objects (see also Section~\ref{subsec:contamination}). Owing to the better angular resolution, HI4PI reveals significantly finer structure in the \ion{H}{i} gas. It also contains fewer image artifacts. For example, in the top left panel LAB appears somewhat blocky, which is due to insufficient spatial sampling (causing aliasing) and in the bottom left panel there are clearly several outlying pixels.

\begin{figure*}[!t]
\centering%
\includegraphics[width=\textwidth,viewport=80 35 1515 760,clip=]{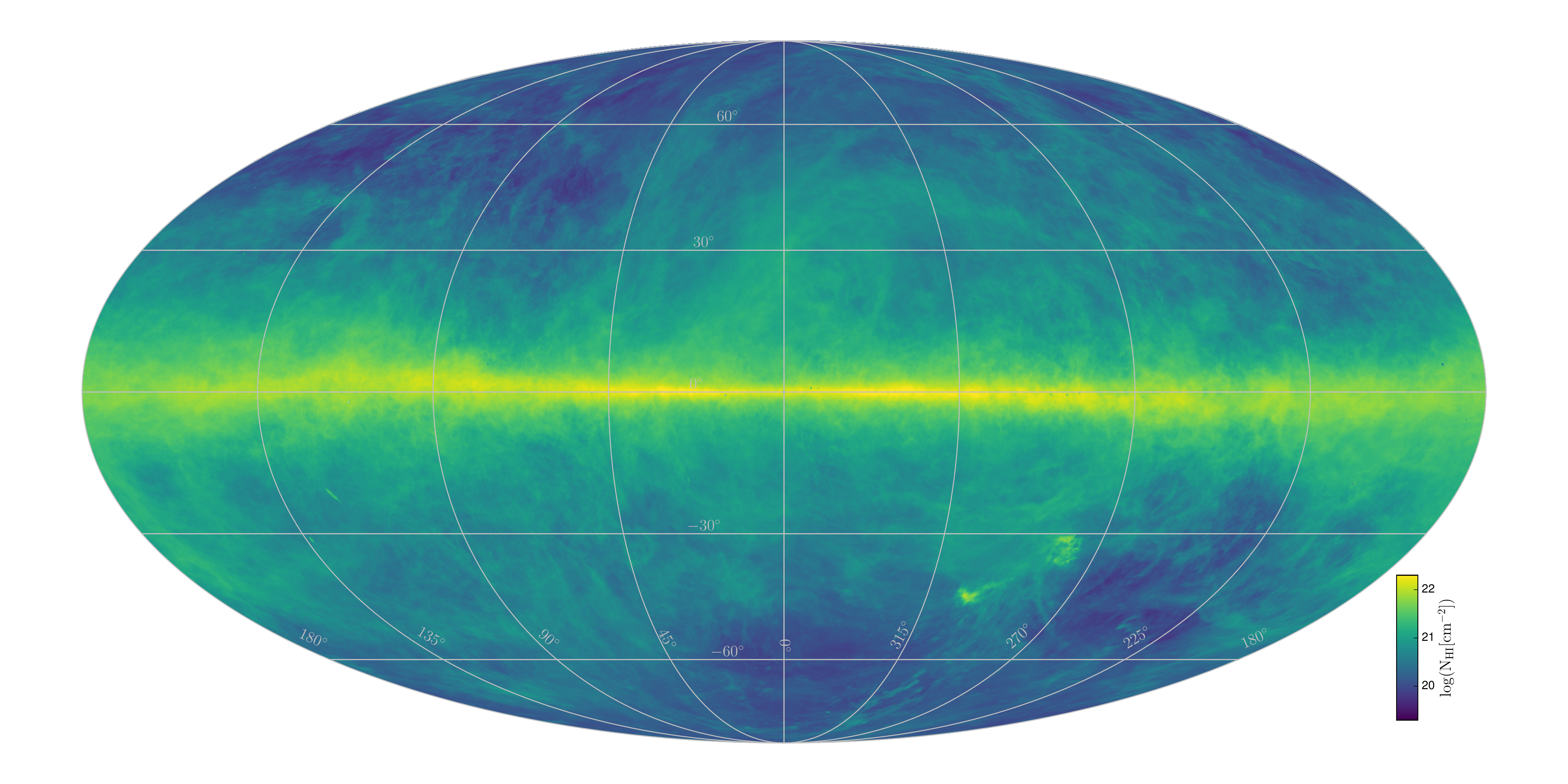}
\caption{HI4PI: all-sky column density map of \ion{H}{i} gas from EBHIS and GASS data as integrated over the full velocity range $-600\leq\varv_\mathrm{lsr}\leq600~\mathrm{km\,s}^{-1}$. The map is in Galactic coordinates using Mollweide projection.}%
\label{fig:nhi_map_mol}%
\end{figure*}

Both surveys, EBHIS and GASS, have a considerable overlap ($-5\degr\lesssim\delta\lesssim0\fdg5$), which allows us to assess the quality of the data by direct comparison. This is presented in detail in Appendix~\ref{sec:ebhis_gass_comparison}, one result being that in some regions in the overlap there are mild differences between EBHIS and GASS. Thus, a mere concatenation of the two surveys (e.g., at $\delta=0\degr$) could lead to discontinuities in the \ion{H}{i} distribution. A simple linear interpolation between EBHIS and GASS, however, is sufficient to mitigate these differences. In the overlap area (limited by $\delta_\mathrm{lo}$ and $\delta_\mathrm{hi}$) we calculate the merged brightness temperature as
\begin{equation}
T_\mathrm{B}^\mathrm{HI4PI}\left(\alpha,\delta\right) = \frac{\delta - \delta_\mathrm{lo}}{\delta_\mathrm{hi} - \delta_\mathrm{lo}}  T_\mathrm{B}^\mathrm{EBHIS}\left(\alpha,\delta\right) + \frac{\delta_\mathrm{hi} - \delta}{\delta_\mathrm{hi} - \delta_\mathrm{lo}}  T_\mathrm{B}^\mathrm{GASS}\left(\alpha,\delta\right)\,,
\end{equation}
and we choose $\delta_\mathrm{lo}=-5\degr$ and $\delta_\mathrm{hi}=0\degr$ as the declination range in which the interpolation is to be applied.

\section{Moment maps}\label{sec:momentmaps}
\subsection{$N_\ion{H}{i}$ map}\label{subsec:nhi}

By integrating spectroscopic data in velocity, one can infer the $N_\ion{H}{i}$ column densities,

\begin{equation}
N_\ion{H}{i}\left[\mathrm{cm}^{-2}\right] = 1.823\cdot10^{18}\int~\mathrm{d}\varv\,T_\mathrm{B}(\varv)\,\left[\mathrm{K\,km\,s}^{-1}\right],\label{eq:coldens}
\end{equation}
where $T_\mathrm{B}(\varv)$ is the brightness temperature profile of the \ion{H}{i} gas \citep[e.g.,][]{wilson13}. The resulting $N_\ion{H}{i}$ map covers the full sky ($4\pi$) and is presented in Fig.~\ref{fig:nhi_map_mol} on a logarithmic intensity scale. In the figure one can see the warp and flaring of the MW disk \citep[][and references therein]{kalberla09}. Furthermore, because of integrating across HI4PI's full velocity range ($-600\leq\varv_\mathrm{lsr}\leq600~\mathrm{km\,s}^{-1}$), the $N_\ion{H}{i}$ map does not only contain MW disk material but also features residing in the MW halo: the intermediate- and high velocity clouds (IVC, HVC) and cloud complexes, as well as extra-galactic objects such as the Magellanic Clouds (\object{LMC} and \object{SMC}) and \object{M\,31}. This contamination usually needs to be considered when working with the $N_\ion{H}{i}$ data and is further discussed in Section~\ref{subsec:contamination}. For completeness, Appendix~\ref{appsec:supp_figures} contains a version of the $N_\ion{H}{i}$ map using a linear intensity scale (Fig.~\ref{fig:nhi_map_mol_linear}).

We note that Eq.~(\ref{eq:coldens}) is only correct in the optically thin limit. For regions of high \ion{H}{i} volume density (usually cold gas, having low spin temperatures), mainly at low Galactic latitudes, self-absorption occurs, such that Eq.~(\ref{eq:coldens}) provides only a lower limit on $N_\ion{H}{i}$ \citep[e.g.,][]{radhakrishnan60,gibson05,braun09,martin15}. \ion{H}{i} absorption spectroscopy can in principle be used to overcome this shortcoming. For example, in the framework of the on-going THOR survey (The \ion{H}{i}, OH, Recombination line survey of the Milky Way), \citet{bihr15} calculate the optical depth toward the giant molecular cloud \object{W\,43} to estimate the true \ion{H}{i} column density. Using Galactic continuum emitters for the \ion{H}{i} absorption technique, however, has the intrinsic issue that only the fraction of \ion{H}{i} gas in front of the continuum emitter can be accounted for. Another approach is to study \ion{H}{i} absorption features toward extra-galactic sources. In both cases, however, one can only derive opacities for a relatively small set of sight lines. Several studies were made to infer a correction factor, $f$, from such samples \citep[see e.g.][and references therein]{dickey00,lee15}, where $f$ is a function of the thin gas-approximated $N_\ion{H}{i}$ value.

Although the aforementioned absorption studies provide very interesting results, it is not yet feasible to perform a proper opacity correction with the full HI4PI data set. Empirically inferred correction factors show too much spread from region to region, preventing application to the full MW disk. The situation may change, once high-resolution absorption-line surveys of the full MW disk are published.

\subsection{Composite maps: combining $N_\ion{H}{i}$ and radial velocities}\label{subsec:composite}

\begin{figure*}[!t]
\centering%
\includegraphics[width=\textwidth,viewport=80 35 1515 760,clip=]{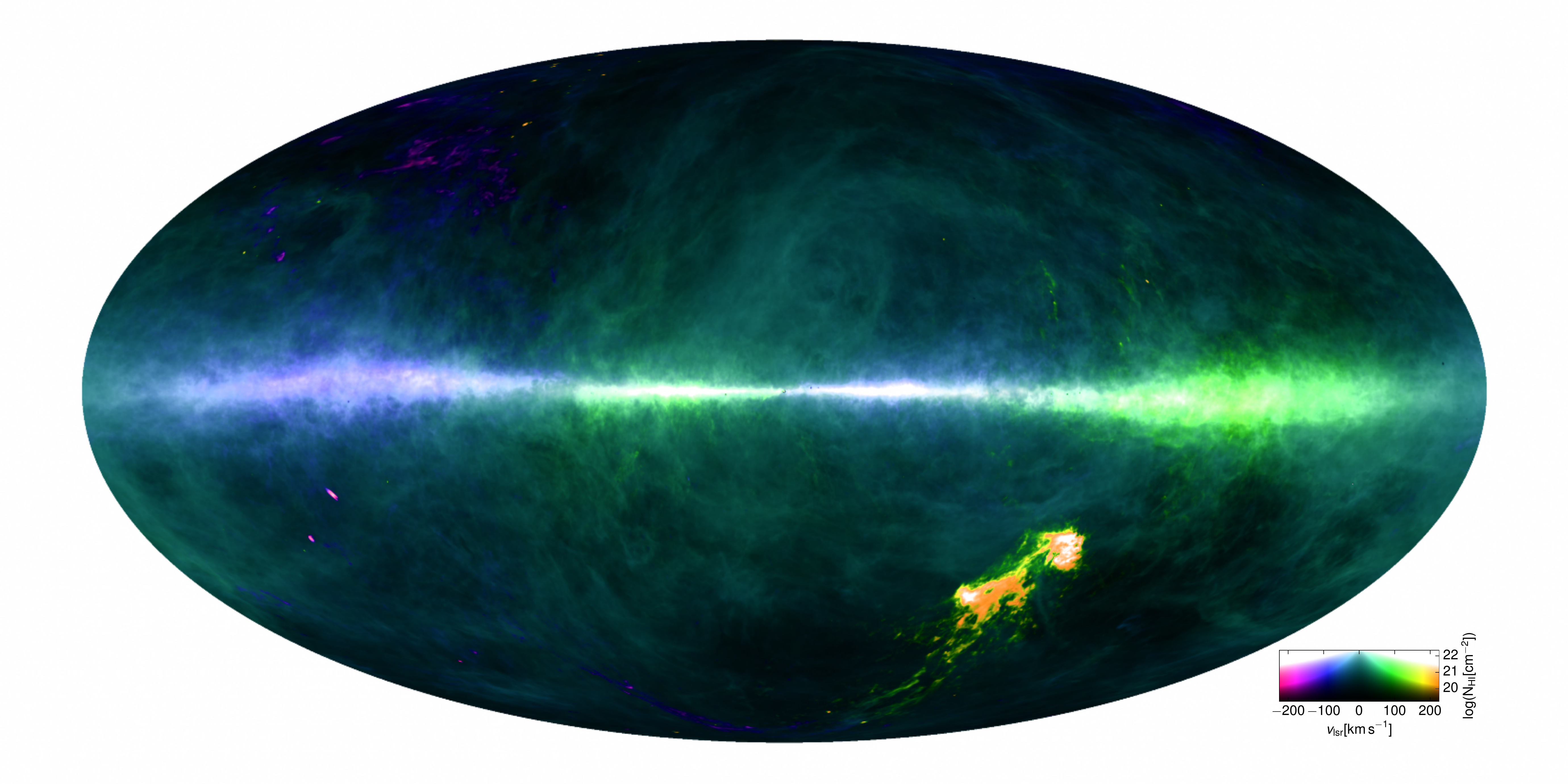}\\[0ex]
\caption{All-sky composite map of the first two moments of the \ion{H}{i} distribution using the same coordinate system and map projection as Fig.~\ref{fig:nhi_map_mol}. The Moment-0 (the column density) determines brightness (log-scale), while the Moment-1 (intensity-weighted radial velocities) is encoded as different hues. The brightness is furthermore modified according to the velocity scale, to render the weaker intermediate- and high-velocity gas more visible. We note that radial velocities in excess of the displayed color scale are subject to clipping: the true radial velocities of red and orange features may be larger than what the colors denote.}%
\label{fig:allsky_composite}%
\end{figure*}

In Fig.~\ref{fig:allsky_composite} we display a composite map of the first two image moments: the $N_\ion{H}{i}$ column density distribution (Moment-0) and the intensity-weighted radial velocity (Moment-1). The former was calculated using Eq.~(\ref{eq:coldens}) from the full spectral dataset (data cube); for the latter a mask was applied to avoid numerical artifacts caused by individual noise peaks in low-column density regions. This mask was constructed by applying a Gaussian filter along the spectral axis followed by a simple thresholding. The combination of both Moment maps was done in the hue--saturation--value (HSV) color space, where Moment-1 is directly used for hue and the (log-scale) $N_\ion{H}{i}$ determines the ``brightness'' (a combination of saturation and value). To visually enhance intermediate- and high velocity cloud features, the brightness was furthermore boosted for larger radial velocities, $\vert \varv_\mathrm{lsr}\vert$. The two-dimensional color wedge in Fig.~\ref{fig:allsky_composite} depicts the transfer function.

A slightly modified version of the composite map is shown in Appendix~\ref{appsec:supp_figures} (Fig.~\ref{fig:allsky_composite_squeezed}) where the chosen radial velocity interval is narrower to enhance the kinematics of the MW disk material. We point out that features with radial velocities beyond the visualized velocity range were not removed from the figure but are assigned one of the two extreme colors (red and orange, respectively). Therefore one has to be careful with the interpretation of the map. For example the Magellanic System has much higher radial velocity than suggested by the color coding. This clipping also occurs in Fig.~\ref{fig:allsky_composite} but to a much lower degree.

Both, Moment-0 and Moment-1, marginalize over velocity and as such a lot of information present in the HI4PI data cubes is lost. This is a severe issue in particular within the Galactic plane where many sight lines usually reveal multiple components with very different radial velocities. As a consequence, all presented Moment maps offer only limited scientific information and one should be careful with the interpretation, especially of the composite maps displayed in Figs.~\ref{fig:allsky_composite} and \ref{fig:allsky_composite_squeezed}.

\subsection{Contamination by extragalactic objects}\label{subsec:contamination}

By providing the HI4PI total column density map we also want to facilitate  cross-correlation studies from the gamma-ray to the far-infrared regime. One aspect of this is that such observations need to be corrected for foreground absorption caused by neutral hydrogen, or interstellar matter quantitatively traced by \ion{H}{i}. Toward the high Galactic latitude sky, significant absorption by \ion{H}{i} is not commonly expected. Thus, one could be inclined to consider the total column density map, integrated in radial velocity over $\vert \varv_\mathrm{lsr}\vert \leq 600\,\mathrm{km\,s}^{-1}$ ($470\,\mathrm{km\,s}^{-1}$ for GASS; compare Section~\ref{sec:dataproducts}), as a quantitative tracer of the Milky Way's interstellar matter. But this expectation is not entirely correct: the HyperLEDA database\footnote{http://leda.univ-lyon1.fr/} \citep{makarov14} contains about 5\,800 entries (status: June 2016) with \ion{H}{i} flux densities above $1~\mathrm{Jy\,km\,s}^{-1}$ populating this radial velocity regime, $\sim$500 of which are most likely associated with galaxies. Only a fraction is connected to very bright \ion{H}{i} emission: the number of objects drops to about 600 (87 galaxies) if the flux density threshold is increased to $50~\mathrm{Jy\,km\,s}^{-1}$. Nevertheless, one should consider the contamination of the HI4PI column density map with non-Milky Way objects.

In addition, there are intermediate-velocity clouds \citep{wakker01} and high-velocity clouds \citep{wakker97} in the halo of the Milky Way (which account for the non-galaxy objects in the HyperLEDA catalog). In Fig.~\ref{fig:allsky_composite}, IVC and HVC complexes, as well as several galaxies can be easily identified by eye. IVCs are thought to populate the inner halo. In the FIR, it has been found that the dust opacity and the dust temperature of IVCs differ significantly from local gas properties \citep[][their Fig. 18]{planck11}. However, they have similar molecular content, densities, and dust-to-gas as the MW disk gas \citep[see e.g.][]{wakker04,roehser16b}. Therefore, it is usually, but not always, appropriate to include the IVC sky into the \ion{H}{i} column density when analyzing absorption processes (of external radiation), while other galaxies and HVCs often have to be treated distinctly. In the following, we will discuss these two classes of objects and their impact in more detail.

High-velocity clouds cover a significant fraction of the \ion{H}{i} sky \citep[see][their Fig.~1]{putman12}. HVCs differ in chemical composition from IVC and Milky Way gas \citep{gibson01} and appear to have insignificant amounts of dust \citep{wakker86,bates88,boulanger96,planck11,saul14,lenz16} and molecular gas \citep{richter01}. For soft X-rays $E\leq0.5~\mathrm{keV}$ the photoelectric attenuation is mainly caused by hydrogen and helium \citep[e.g.,][]{snowden90,herbstmeier94,herbstmeier95,wilms00}. Thus the HI4PI column density map accounts quantitatively very well for the Galactic photoelectric absorption of redshifted X-ray sources. For higher X-ray photon energies, however, the metallicity of the interstellar gas needs to be taken into account. Here, using HI4PI $N_\ion{H}{i}$ as a unique measure for the X-ray attenuation would be an oversimplification. For a proper treatment of the photoelectric absorption cross section one needs to weight the different gas columns and their metallicity properly. Therefore, we recommend downloading HI4PI data cubes from the archive in order to evaluate the gas columns for the Milky Way and HVC velocity regimes separately.
As an example, in Fig.~\ref{fig:hi4pi_lab_comparison} (row~1) the Draco cloud \citep{goerigk83,mebold85,rohlfs89,miville16} is shown, which adds considerably to the amount of $N_\ion{H}{i}$ observed.

Extragalactic objects in the HI4PI database are all local galaxies. Prominent examples are the Magellanic Clouds, \object{M\,31}, \object{M\,33} as well as the \object{M\,81}/\object{M\,82} and Sculptor galaxy groups \citep[e.g.,][]{yun94,kerp16a}. All these extragalactic objects contribute significantly to, or even dominate, the HI4PI total column density map at certain positions. While more compact galaxies are easily identified, the contamination by extended objects such as the Magellanic Clouds often demands a more detailed investigation. Furthermore, the \ion{H}{i} profile of many galaxies overlaps in velocity with MW disk material. As an example, Fig.~\ref{fig:hi4pi_lab_comparison} displays a part of the Leading Arm region \citep[row~2;][]{bruens05,venzmer12, for13} of the Magellanic Clouds \citep[row~3;][]{kim03,bruens05} and the M31/M33 region \citep[row~4;][]{braun03,thilker04,wolfe16,kerp16a}.

In conclusion, we emphasize that the published HI4PI $N_\ion{H}{i}$ map is not free of contamination with extra-galactic objects and HVCs. It may be possible to remove a majority of them from the data with some extra work. However, for objects that overlap in velocity with MW disk material this task gets challenging and more error-prone. Therefore, we defer this topic to a future paper. In the meantime, we recommend using the full spectral data toward problematic areas of interest (to distinguish between the different components) and making use of supplementary information where available.

\section{Data products}\label{sec:dataproducts}

The individual GASS and EBHIS datasets are already available to the scientific community. GASS spectral profiles and data cubes can be accessed via the Bonn \ion{H}{i}-survey server\footnote{\url{https://www.astro.uni-bonn.de/hisurvey/}}. The data cubes, however, are limited in size to $10\degr\times10\degr$ to minimize the computing effort and reduce the traffic on the server. EBHIS profiles are also available on the \ion{H}{i}-survey server. Furthermore pre-computed data cubes of size $20\degr\times20\degr$ and all-sky column density maps of EBHIS are available on CDS\footnote{\url{http://cdsweb.u-strasbg.fr/cgi-bin/qcat?J/A+A/585/A41}}.

The new data products are made available on CDS, in a similar manner as previously for EBHIS, but for the Galactic instead of the Equatorial coordinate system. We note that for certain science cases the EBHIS and GASS databases may still be better suited: the original EBHIS data offer higher angular resolution and GASS has better spectral resolution than what is now published in HI4PI.

We provide \ion{H}{i} column density maps as FITS images in various map projections \citep[according to WCS;][]{greisen02,calabretta02} such as Mollweide (MOL) and Hammer-Aithoff (AIT). The \ion{H}{i} column density map in Fig.~\ref{fig:nhi_map_mol} is based on the Mollweide projection, for example. Furthermore, a version of the $N_\ion{H}{i}$ distribution on the HEALPix grid \citep{gorski05} is available. HEALPix is widely used for all-sky datasets. For this we chose an nside parameter of 1024, which leads to a pixel size of $3\farcm44$.

For the column density maps, we chose to integrate over the full velocity range of each of the two surveys (EBHIS: $\vert \varv_\mathrm{lsr}\vert \leq 600\,\mathrm{km\,s}^{-1}$, GASS: $470\,\mathrm{km\,s}^{-1}$). One could have used the smaller interval (GASS) also for the northern hemisphere, but then several features (e.g., \object{M\,31} and \object{M\,33}, see Fig.~\ref{fig:hi4pi_lab_comparison} bottom panel) would have been affected.

Spectral data are presented in two flavors: (1) FITS images (data cubes), again in various projections and (2) as FITS binary table containing spectra on the HEALPix grid. Because a full-sky HI4PI data cube has a size of about 25~GB, we also offer smaller $20\degr\times20\degr$ subcubes, which in many cases should be more convenient to work with. For the same reason, the HEALPix binary table is split into smaller files, each containing the spectra for one of the 192 HEALPix pixels of the nside$=$4 representation.

\section{Summary}\label{sec:summary}

We have presented the new HI4PI all-sky \ion{H}{i} survey, that we constructed from the recent EBHIS and GASS data sets. It is freely available to the scientific community via CDS. HI4PI supersedes the LAB survey for any astronomical study that needs accurate \ion{H}{i} profiles or column density values of the Milky Way. The various formats in which we release the data sets will make it easy to use HI4PI as a drop-in replacement for the LAB survey: (1) all-sky column density maps as FITS images and HEALPix-grid binary FITS tables, (2) all-sky FITS data cubes, and subcubes of size $20\degr\times20\degr$, and (3) spectral profiles in HEALPix-grid binary FITS tables. All of the FITS images (data cubes) are provided in various WCS projections for the Galactic coordinate system.

The high data quality of the HI4PI constituents, EBHIS and GASS, has already been demonstrated \citep{kalberla15,winkel16a}. Both are corrected for stray radiation. The joint database excels in angular resolution and sensitivity compared to any previous all--sky \ion{H}{i} survey of the Milky Way.

\begin{acknowledgements}

We are grateful to Alexander Kraus for carefully proofreading the manuscript and for his valuable comments. Furthermore, we like to thank the referee, Peter Martin, for his suggestions, which improved the overall presentation of the paper significantly.

EBHIS is based on observations with the 100-m telescope of the MPIfR (Max-Planck-Institut für Radioastronomie) at Effelsberg. The Parkes Radio Telescope is part of the Australia Telescope which is funded by the Commonwealth of Australia for operation as a National Facility managed by CSIRO. The authors thank the Deutsche Forschungsgemeinschaft (DFG) for support under grant numbers KA1265/5-1, KA1265/5-2, KE757/7-1, KE757/7-2, KE757/7-3, and KE757/11-1. BW was partially funded by the International Max Planck Research School for Astronomy and Astrophysics at the Universities of Bonn and Cologne (IMPRS Bonn/Cologne). LF was also a member of IMPRS Bonn/Cologne. DL is a member of the Bonn--Cologne Graduate School of Physics and Astronomy (BCGS). UH acknowledges the support by the Estonian Research Council grant IUT26-2, and by the European Regional Development Fund (TK133). NM-G is supported by Australian Research Council Future Fellowship, FT150100024. JB is also an Adjunct Astronomer at the National Radio Astronomy Observatory. DJP was partially supported by NSF CAREER grant AST-1149491.

This research has made use of NASA's Astrophysics Data System. This research has made use of the NASA/IPAC Extragalactic Database (NED) which is operated by the Jet Propulsion Laboratory, California Institute of Technology, under contract with the National Aeronautics and Space Administration. This research has made use of the SIMBAD database, operated at CDS, Strasbourg, France.

We would like to express our gratitude to the developers of the many C/C++ and Python libraries, made available as open-source software, which we have used: most importantly, NumPy \citep{NumPy} and SciPy \citep{SciPy}, Cython \citep{Cython}, and Astropy \citep{Astropy}. Figures have been prepared using matplotlib \citep{Matplotlib} and in part using the Kapteyn package \citep{KapteynPackage}.

\end{acknowledgements}

\bibliographystyle{aa} 
\bibliography{references} 

\appendix
\section{Comparing EBHIS and GASS}\label{sec:ebhis_gass_comparison}

The two surveys, EBHIS and GASS share a relatively large area on the sky ($\sim$1800~deg$^2$ or $4.4\%$). This gives us the opportunity to directly compare both datasets, with the aim of assessing the data quality in the HI4PI survey. For single-dish \ion{H}{i} data sets there are typically four major effects to be considered: (1) flux-density (or brightness-temperature) calibration uncertainties, (2) accuracy of the radial velocity scale, (3) baseline problems caused by uncertainties in the determination of the instrumental bandpass and system temperature curves but also from residual errors in the stray--radiation correction, and (4) residual radio frequency interference not identified by the RFI flagger or improperly removed after detection. These aspects have been discussed in detail previously \citep{mcclure09,kalberla10,kalberla15,winkel16a}. For convenience, we will briefly repeat the most important findings here and complement these with additional analyses with a focus on comparing the overlap region of EBHIS and GASS.

To study the common survey area, a data cube was produced for each of EBHIS and GASS covering $-4\fdg5\leq\delta\leq0\fdg5$ using exactly the same angular and spectral grid for both surveys, smoothed to the HI4PI resolution. This gives us the possibility to compare $N_\ion{H}{i}$ (by integrating the data cube along the spectral axis) and position--velocity slices that allow a better visualization of certain problems with the data compared to position--position diagrams.

In Fig.~\ref{fig:nhi_map_diff1} the resulting $N_\ion{H}{i}$ values are compared for a large portion of the overlap region. Each of the four stripes contains a 4-hour interval in right ascension. The upper panels of each stripe display the GASS column density, $N_\ion{H}{i}^\mathrm{GASS}$, the lower panels the relative difference, $(N_\ion{H}{i}^\mathrm{EBHIS} - N_\ion{H}{i}^\mathrm{GASS})/N_\ion{H}{i}^\mathrm{EBHIS}$ (in percent). In the difference plots significant deviations of up to $10-15\%$ are revealed toward some regions. Furthermore, a rectangular pattern is prominently visible, which we can attribute to the GASS. In the GASS column density maps, it is hardly visible, though. The pattern is related to the GASS scanning strategy, which is made up from the two orthogonal scan directions along right ascension and declination. The stripes are likely caused by correlator or receiver failures \citep[see][their Section~2.7.]{kalberla15}. Only the most severe degradations could be eliminated, low level scanning problems remained.

Some of the other $N_\ion{H}{i}$ differences are related to the $5\degr\times5\degr$ survey fields of EBHIS (marked with small black arrows in Fig.~\ref{fig:nhi_map_diff1}). Far-side-lobe SR contamination is well known to produce patches that are related to the individual observing sessions \citep[compare][]{kalberla05,kalberla10,winkel16a}. These SR contributions are typically at a level below $40~\mathrm{mK}$, occasionally up to $100~\mathrm{mK}$. This is close to the average noise level of $\sim43~\mathrm{mK}$ in HI4PI. So far, we were unable to identify the origin of the remaining SR features, which would be necessary to improve the SR correction software. We also note that SR contamination would be most severe in low-column density regions, but it appears that at higher Galactic latitudes the residual contamination is rather insignificant \citep[compare][]{martin15}.

More worrying, however, is that also other medium-scale structures (several degrees but smaller than the $5\degr\times5\degr$ fields) appear in the difference maps, which are neither clearly correlated with the observed column density nor with the scan strategy. At the moment, it is not even clear which of the two surveys is responsible for these structures. Again, the second sky coverage of EBHIS will potentially shed some light onto this issue.

To visualize the baseline quality and other effects, position--velocity (p--v) diagrams are well-suited. Figs.~\ref{fig:ebhis_gass_pv1} and \ref{fig:ebhis_gass_pv2} display the difference between EBHIS and GASS (top left panel), the GASS data with full intensity scale (top right panel), and EBHIS and GASS data with a zoomed-in intensity scale (bottom row), which allows us to trace faint brightness temperature features.

As mentioned above, EBHIS data were smoothed to the GASS angular resolution, such that both surveys have comparable \ion{H}{i} feature sizes and brightness temperature noise. However, the different gridding kernel size leads to a different small-scale correlation in the noise, an effect which is explained in detail in \citet[][their Appendix~1]{winkel16b}. It can be seen in Figs.~\ref{fig:ebhis_gass_pv1} and \ref{fig:ebhis_gass_pv2}: noise grains in GASS appear smaller. However, this should not have any negative effect on analyses using the data.

In the following, we will discuss the HI4PI data quality in more detail based on the three Figs.~\ref{fig:nhi_map_diff1} to \ref{fig:ebhis_gass_pv2} and previous work.

\afterpage{
\begin{landscape}
\begin{figure}[!p]
\centering%
\includegraphics[width=1.3\textwidth,viewport=14 15 1437 267,clip=]{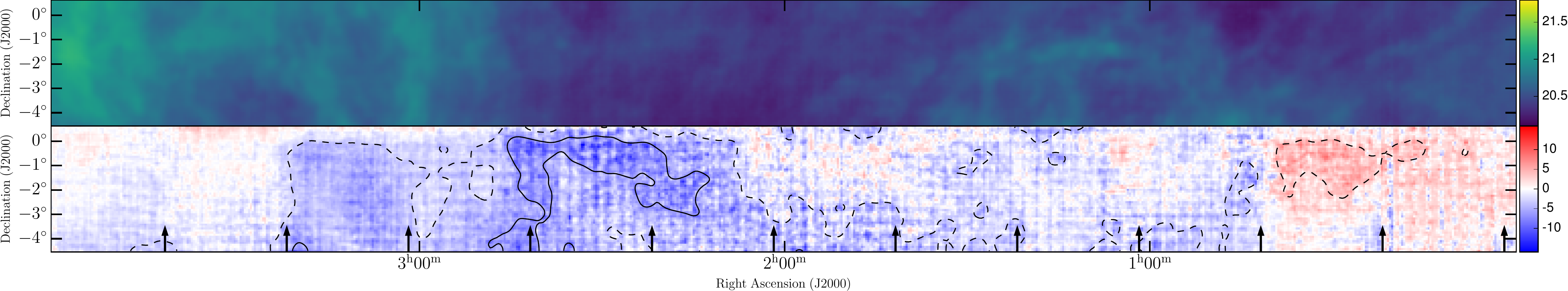}\\[0ex]
\includegraphics[width=1.3\textwidth,viewport=14 15 1437 267,clip=]{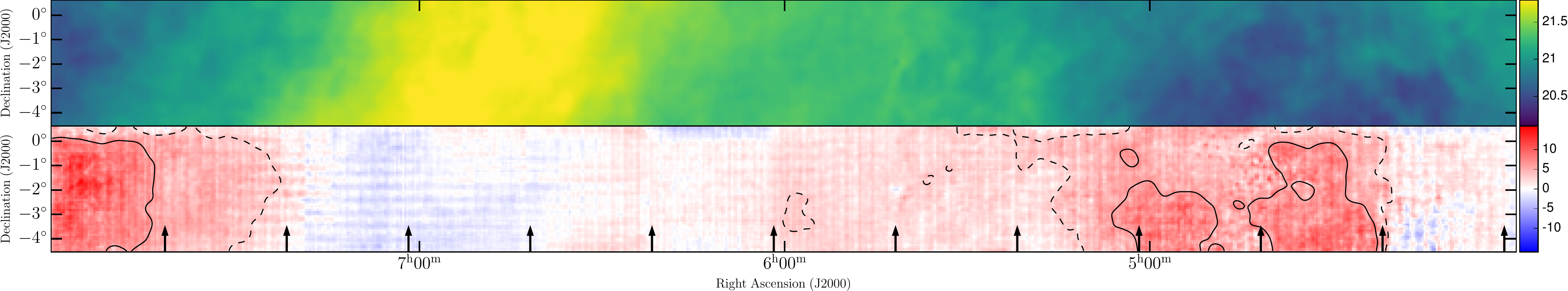}\\[0ex]
\includegraphics[width=1.3\textwidth,viewport=14 15 1437 267,clip=]{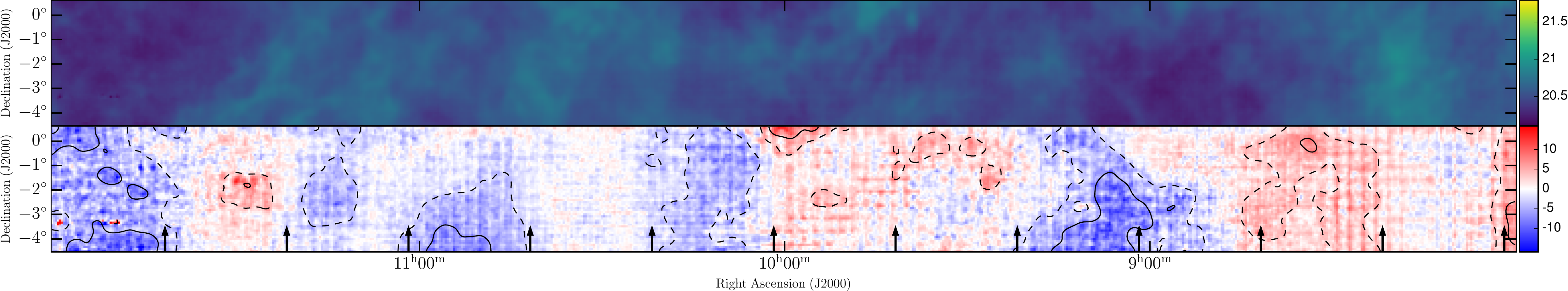}\\[0ex]
\includegraphics[width=1.3\textwidth,viewport=14 15 1437 267,clip=]{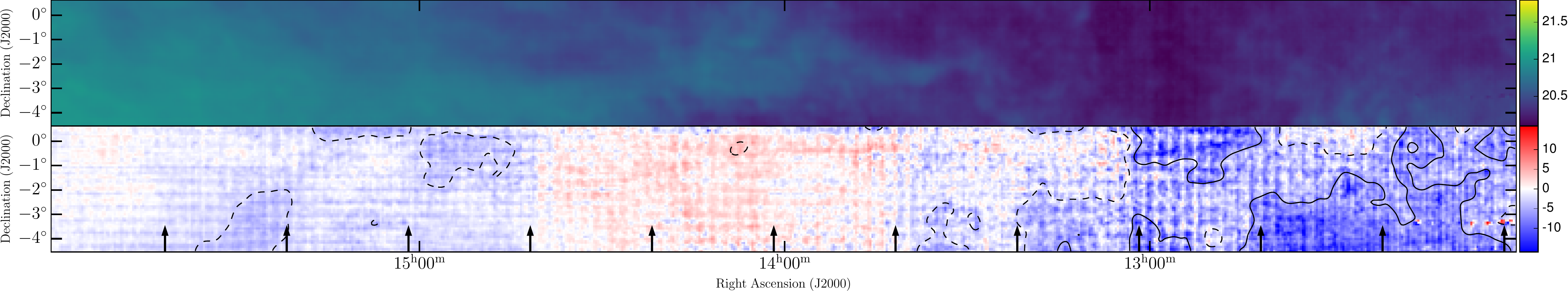}
\caption{EBHIS vs. GASS column densities in the right ascension range $0^\mathrm{h}$ to $16^\mathrm{h}$. The upper panel in each stripe shows $\log(N_\ion{H}{i}^\mathrm{GASS} [\mathrm{cm}^{-2}])$, the lower panels show the relative difference $(N_\ion{H}{i}^\mathrm{EBHIS} - N_\ion{H}{i}^\mathrm{GASS})/N_\ion{H}{i}^\mathrm{EBHIS}$ in percent. For convenience, each of the lower panels also contains contour levels for $\pm3\%$ (dashed lines) and $\pm6\%$ (solid lines), as well as black arrows to mark the EBHIS field limits.}%
\label{fig:nhi_map_diff1}%
\end{figure}
\end{landscape}
}
\afterpage{
\begin{landscape}
\begin{figure}[!p]
\centering%
\includegraphics[width=1.3\textheight,clip=]{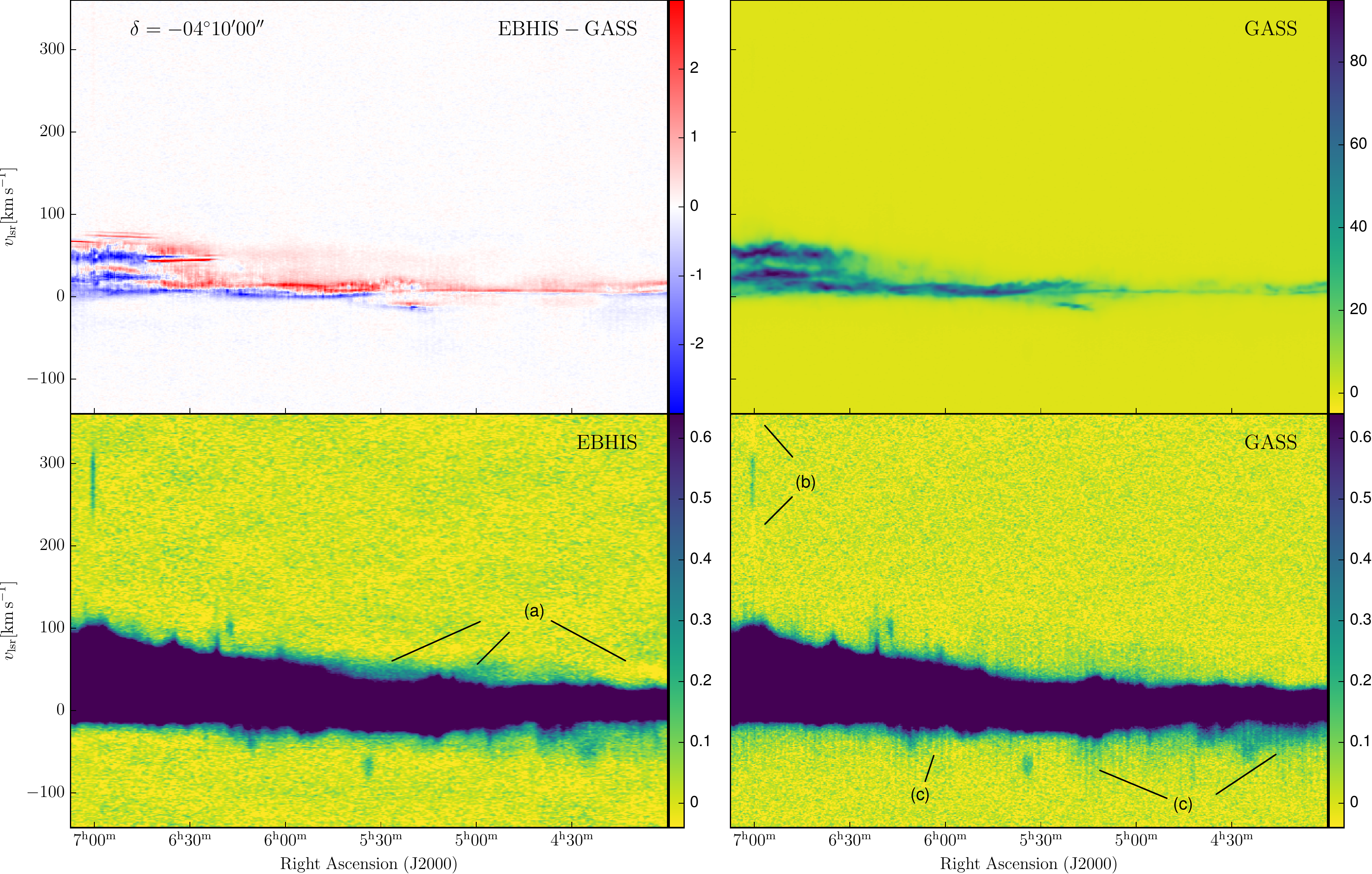}
\caption{Typical EBHIS and GASS baseline problems. The four panels show a position--velocity slice of the overlap-region datacube. \textit{Top left panel:} difference between EBHIS and GASS, \textit{top right panel:} GASS data, and \textit{bottom panels:} EBHIS and GASS data with a zoomed-in intensity scale. The labeled features are: (a) larger-angular scale baseline fluctuations in direct vicinity of the Milky Way disk \ion{H}{i} profile in EBHIS data; (b) baseline deficiencies around non-disk objects in GASS, caused by improper flagging; and (c) increase in baseline level associated to the rectangular pattern visible in Fig~\ref{fig:nhi_map_diff1}. All colorbars show brightness temperature, $T_\mathrm{B}$, in units of Kelvin. See text for further discussion.}%
\label{fig:ebhis_gass_pv1}%
\end{figure}
\end{landscape}
}

\afterpage{
\begin{landscape}
\begin{figure}
\centering%
\includegraphics[width=1.3\textheight,clip=]{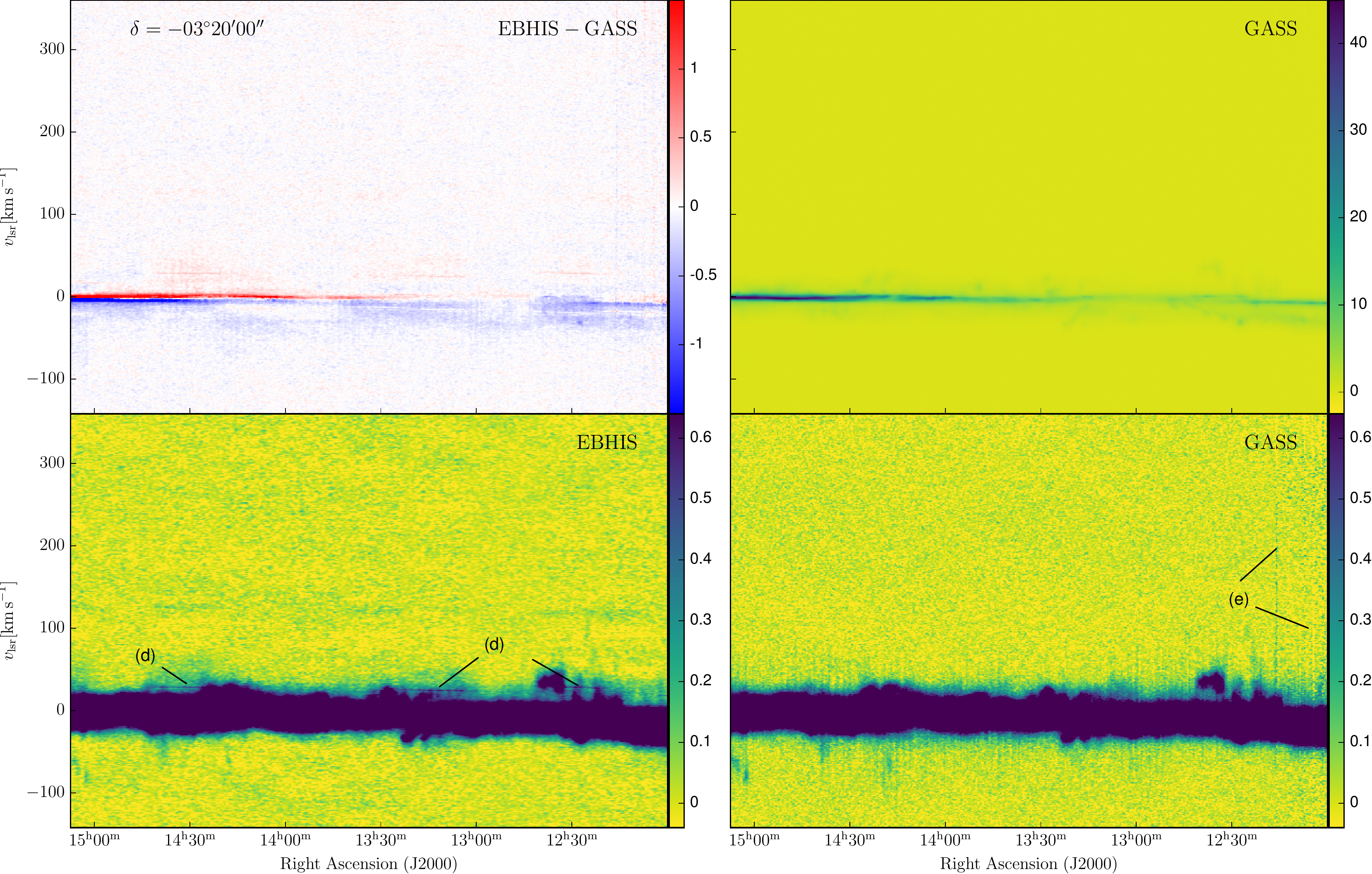}
\caption{As Fig.~\ref{fig:ebhis_gass_pv1}, showing typical EBHIS and GASS residual RFI features. The labeled features are: (d) narrowband RFI within or very close to the Milky Way disk emission in EBHIS data; (e) rare broadband events, not detected by the GASS RFI flagger. See text for further discussion.}%
\label{fig:ebhis_gass_pv2}%
\end{figure}
\end{landscape}
}

\subsection{Brightness temperature calibration}\label{subsec:intensitycalibration}

\citet{winkel16a} demonstrated the overall very consistent brightness temperature calibration of EBHIS with respect to the established LAB survey data, by comparing $N_\ion{H}{i}$ and $T_\mathrm{B}$ pixel-wise. They report on a nearly perfect one-to-one correlation for both quantities
\begin{align}
N_\ion{H}{i}^\mathrm{EBHIS} &= 1.0023(4) \cdot N_\ion{H}{i}^\mathrm{LAB} - 0.3(3) \cdot10^{18}~\mathrm{cm}^{-2}\,,\\
T_\mathrm{B}^\mathrm{EBHIS} &= 1.0000(6) \cdot T_\mathrm{B}^\mathrm{LAB} + 12.2(5)~\mathrm{mK}\,.
\end{align}
Similar results had previously been obtained by \citet{kalberla15} for GASS and LAB\footnote{After re-calibrating GASS to correct for an inconsistent intensity scale between LDS and IAR survey; see \citet{kalberla15} and \citet{winkel16a} for details.}. \citet{martin15} compared selected small-area regions at high Galactic latitudes observed by EBHIS, GASS, and the Green Bank Telescope and also find a consistent calibration scale between the three data sets. Furthermore, they discuss the SR correction quality for the low-column density regime in detail.

Another noteworthy analysis presented by \citet{winkel16a} is the study of typical systematic uncertainties of the $N_\ion{H}{i}$ and $T_\mathrm{B}$ ensemble distributions. For this, the authors compared the relative difference between EBHIS and LAB column densities and brightness temperatures as a function of $N_\ion{H}{i}$ and $T_\mathrm{B}$, respectively. They could show that the ensemble scatter can be explained by thermal noise plus a 2.5\% contribution, which they attribute to intensity calibration errors; see their Section~5.2. We note that the thermal noise is a function of radial velocity and usually higher in regions of high brightness temperature. The resulting uncertainty in $N_\ion{H}{i}$ is a strong function of $N_\ion{H}{i}$ itself \citep[see][their Fig.~15]{winkel16a}. For low column densities ($\log N_\ion{H}{i} \lesssim20.25$) the $1\sigma$ (68\%) confidence level is about 6\%, while for high column densities ($\log N_\ion{H}{i} \gtrsim21.5$) the uncertainty is below 2.5$-$3\%.

Therefore, the seemingly large deviations in the difference maps in Fig.~\ref{fig:nhi_map_diff1} are still consistent with the findings presented in \citet{winkel16a}. Most of the high-column-density regions in the overlap area show less than 3\% deviation. Even in the lower column density regions the deviations rarely exceed 6\%. To guide the eye, we have marked the $\pm3\%$ and $\pm6\%$ levels with contours in Fig.~\ref{fig:nhi_map_diff1}. To reduce clutter due to the very small-scale rectangular pattern, we spatially smoothed the differences prior to contour computation.

\subsection{Radial velocities}\label{subsec:vrad}

A striking feature in Figs.~\ref{fig:ebhis_gass_pv1} and \ref{fig:ebhis_gass_pv2} are $S$-shaped artifacts in the difference spectra (top left panels). Today, we attribute this to a marginal radial velocity mismatch between EBHIS and GASS of about $\Delta\varv \approx -0.35~\mathrm{km\,s}^{-1}$, previously reported in \citet[][see their Fig.~A.3]{winkel16a}. When calculating the difference of the two data sets, a slight velocity shift will effectively act like a numerical derivative operator, which enhances the steepest gradients of each \ion{H}{i} profile. Often these are located in the wings of the bright  \ion{H}{i} peaks, for example close to $\varv_\mathrm{lsr}\approx0~\mathrm{km\,s}^{-1}$ in Fig.~\ref{fig:ebhis_gass_pv2}. In Fig.~\ref{fig:ebhis_gass_pv1} the \ion{H}{i} profile is more complex, leading to a more complicated residual pattern in the difference panel. We could not identify the cause for the slight velocity shift so far. \citet{winkel16a} find an insignificant shift between LAB and EBHIS, while the shift between LAB and GASS is similar to the shift between EBHIS and GASS. This makes it somewhat more likely that the GASS data set is introducing the problem, and not EBHIS.

\subsection{Baseline quality}\label{subsec:baselines}

As discussed in Section~\ref{sec:observations}, EBHIS and GASS use very different strategies to model the baselines. EBHIS uses 2-D polynomial fitting of the time--frequency plane, while GASS does 1-D polynomial fitting in each individual spectral dump. In both cases, \ion{H}{i} emission (and absorption) needs to be identified and masked prior to the fit, otherwise the resulting baseline model would be strongly biased. The parametrization of these masks is implemented as an iterative process in both data reduction pipelines. A polynomial fit is performed, after which outliers in the residual spectrum (data minus fit) are searched for and flagged. This process is repeated until the solution has converged \citep[for details we refer to][]{mcclure09,kalberla10,winkel16a}. Prior information, if available, can also help to improve the baseline solution, especially during the very first iteration. Both surveys used LAB data to define priors. In addition, EBHIS utilized the HyperLEDA database \citep{makarov14} to mask extragalactic \ion{H}{i} objects and NVSS \citep{condon98} to flag strong continuum sources ($\geq$1.5~Jy). Because several \ion{H}{i} galaxies have radial velocities falling into the MW velocity regime, using the HyperLEDA database is beneficial even for the MW velocity regime. This can be seen in Fig.~\ref{fig:ebhis_gass_pv1} (label b), where a baseline deficiency is visible around the galaxy pair \object{HIZSS003A/B} \citep{begum05} in the GASS data.

When interpolating across any masked feature, the polynomial fitting becomes uncertain, in particular across the very broad velocity interval in which \ion{H}{i} emission from the MW disk is present. This gives rise to the features (label a) in Fig.~\ref{fig:ebhis_gass_pv1}. In EBHIS, with the 2-D approach, such baseline defects are usually stretched over several degrees, while in GASS baseline fitting artifacts are expected to be more localized features, as each input profile is individually processed.

The regular rectangular pattern discussed in Appendix~\ref{subsec:intensitycalibration} is also visible in Fig.~\ref{fig:ebhis_gass_pv1}, annotated with label (c). Technically, it is not caused by baseline fitting, but it produces features in the data mimicking typical baseline-fitting artifacts.

As each receiver feed's signal has to be processed individually before gridding, the baseline fitting must be applied to spectra noisier than in the final data cube. This is a problem for EBHIS, because the structures in the raw-data baselines are a function of the incident continuum irradiation \citep[and therefore of time, see][]{winkel16a}. In Fig.~\ref{fig:ebhis_gass_pv1} one can see that the underlying baseline (noise floor) in EBHIS appears somewhat less flat than in GASS.

Unrelated to the baseline fitting itself, but causing similar features, is an improper SR correction. Baseline fitting errors and SR contributions have different origins and different characteristic shapes. A good fraction of the errors is found in distinct radial velocity intervals. Accordingly the corrections are independent from each other. However remaining problems on a low level and in velocity close to the MW emission line may be hardly distinguishable. At this point objective criteria for further improvements of the corrections are missing and it is not possible to increase the quality of the data without additional information such as from a second independent sky coverage. Feature (a) in Fig.~\ref{fig:ebhis_gass_pv1}, as an example, could be due to baseline uncertainties but may also be caused by improper SR correction. For a thorough discussion of SR correction in both surveys we refer to \citet{kalberla10} and \citet{winkel16a}.

\setcounter{section}{2}
\setcounter{figure}{0}

\begin{figure*}[!t]
\centering%
\includegraphics[width=\textwidth,viewport=80 35 1515 760,clip=]{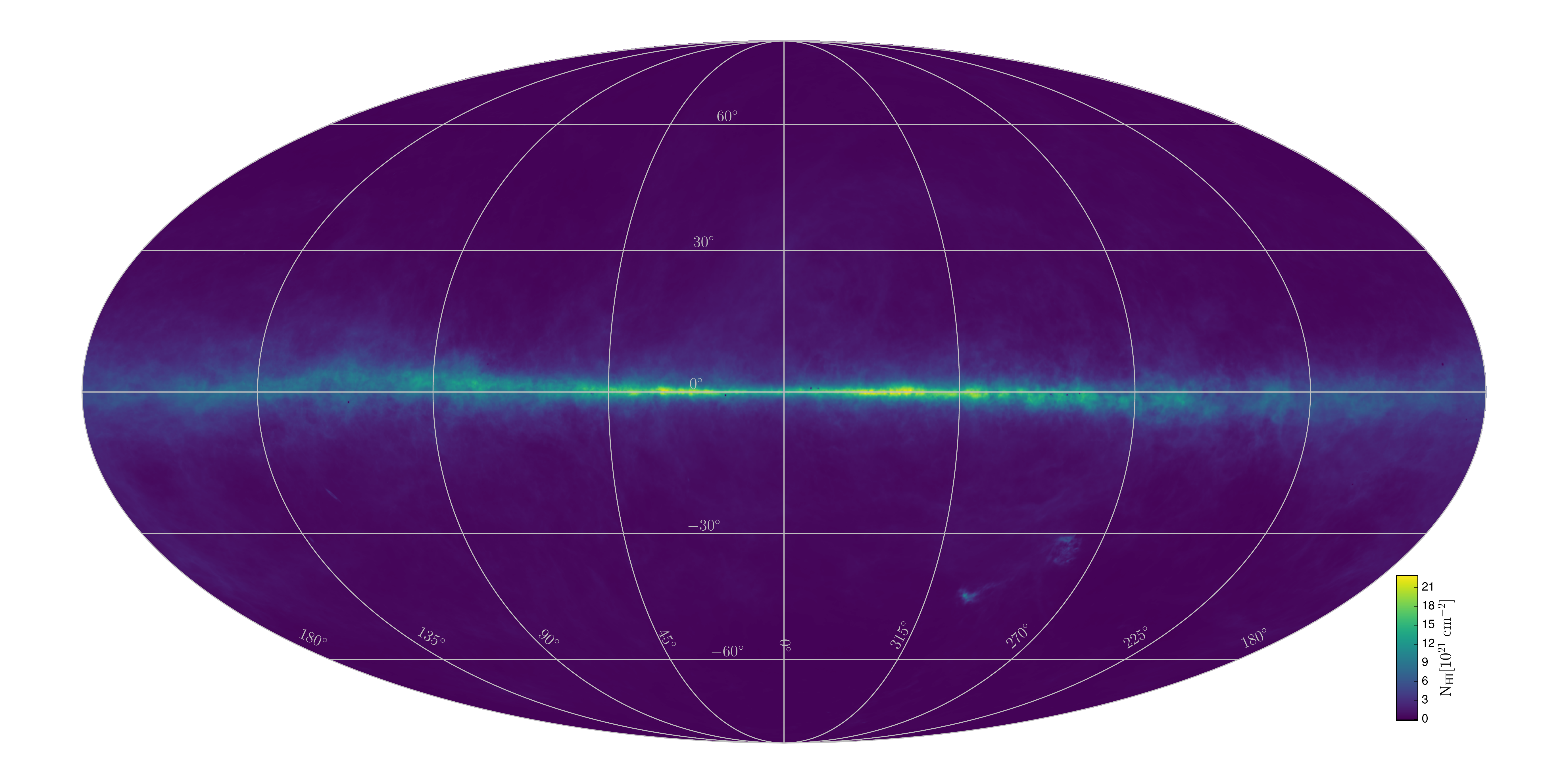}
\caption{As Fig.~\ref{fig:nhi_map_mol} but with linear intensity scale.}%
\label{fig:nhi_map_mol_linear}%
\end{figure*}

\begin{figure*}[!t]
\centering%
\includegraphics[width=\textwidth,viewport=80 35 1515 760,clip=]{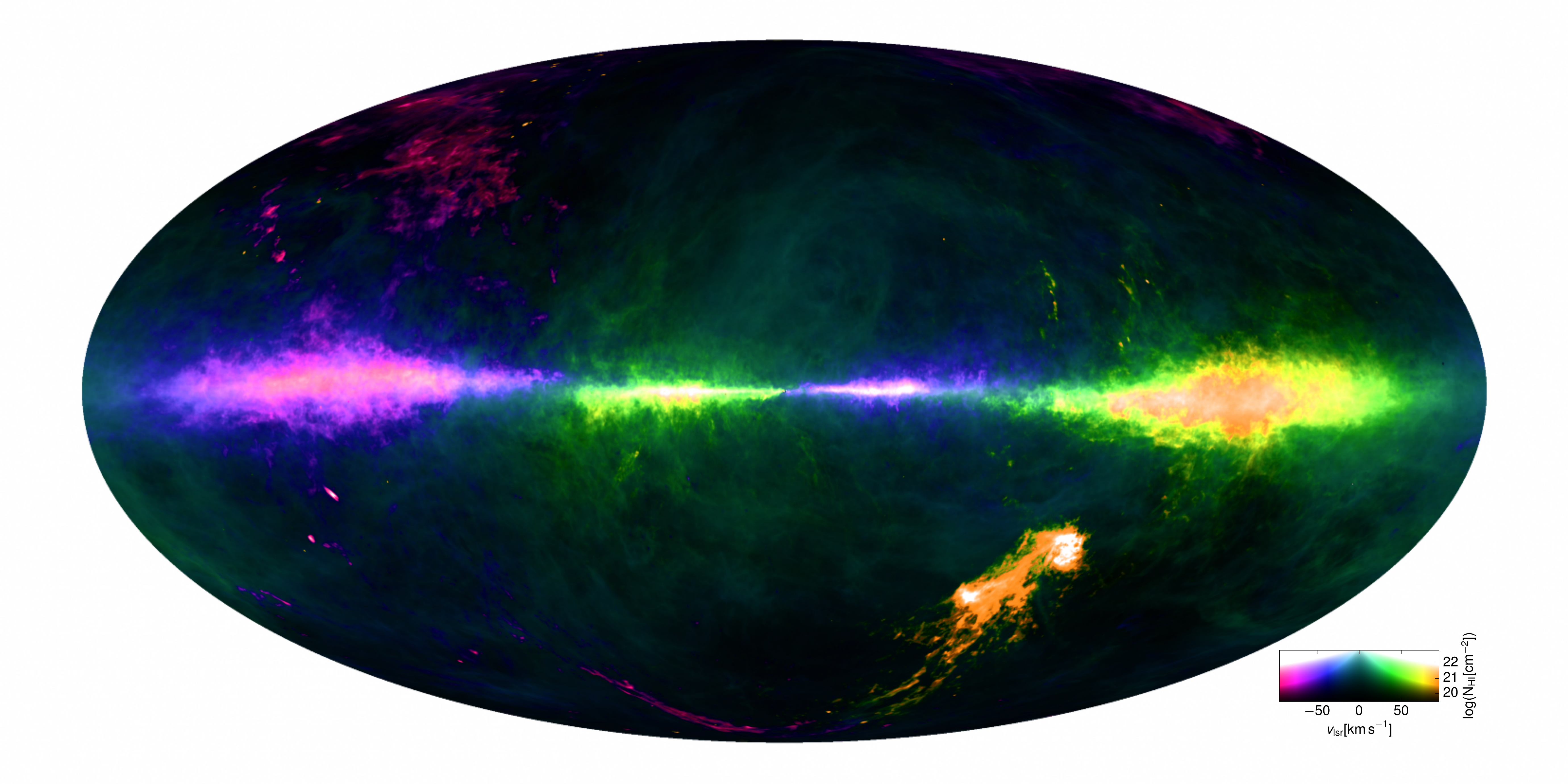}\\[0ex]
\caption{As Fig.~\ref{fig:allsky_composite} displaying a narrower velocity interval to enhance kinematic features in the Milky Way disk. Owing to the narrower velocity range, clipping effects are more severe than in Fig.~\ref{fig:allsky_composite}.}%
\label{fig:allsky_composite_squeezed}%
\end{figure*}

\setcounter{section}{1}

\subsection{Residual radio frequency interference features}\label{subsec:rfi}

Figure~\ref{fig:ebhis_gass_pv2} shows another type of artifact in both data sets: residual RFI. For EBHIS, it was reported by \citet{winkel16a} that residual narrow-band RFI exists in the data cubes. As this type of RFI is mostly constant in frequency for a complete observation, usually $5\degr\times5\degr$-areas are affected (the EBHIS observing field size). Close to the brightest parts of the Milky-Way emission, the EBHIS RFI mitigation software had to be used with very conservative threshold levels. Otherwise, narrow Milky Way line emission peaks would sometimes be falsely identified as RFI. As a consequence, some of the narrow-band RFI events within or close to the MW disk emission were not properly removed from the data; see Fig.~\ref{fig:ebhis_gass_pv1} (top left panel, between $6\fh25\leq\alpha\leq7^\mathrm{h}$, $50\leq\varv_\mathrm{lsr}\leq80~\mathrm{km\,s}^{-1}$) and Fig.~\ref{fig:ebhis_gass_pv2} (label d). In other cases, when the RFI was correctly flagged, the removal algorithm failed to subtract the RFI intensity with the appropriate amplitude, such that sometimes spurious features remain in the data. GASS has on average fewer remaining artifacts caused by RFI; an example is visible in Fig.~\ref{fig:ebhis_gass_pv2} (label e). However, the situation for EBHIS will improve once the second sky coverage is available.

\section{Supplementary figures}\label{appsec:supp_figures}


\end{document}